\documentclass[aps,showpacs,graphicx]{revtex4} 
\usepackage{amssymb} 
\usepackage[dvips]{graphicx} 
\usepackage[english]{babel} 
\usepackage{indentfirst} 
\usepackage{amsxtra} 
\usepackage{amsmath} 
\usepackage{supertabular} 
\usepackage{multirow} 
\usepackage [mathcal]{eucal} 

\newcommand{\beq}{\begin{equation}} 
\newcommand{\eeq}{\end{equation}} 
\newcommand{\beqn}{\begin{eqnarray}} 
\newcommand{\eeqn}{\end{eqnarray}} 
\newcommand{\sumint}{\sum \!\!\!\!\!\!\!\!\int } 
\newcommand{\bsigma}{\mbox{\boldmath $\sigma$}} 
\newcommand{\btau}{\mbox{\boldmath $\tau$}}

\newcommand{\pr}{\pi}  
\newcommand{\nt}{\nu}  
\newcommand{\bn}{\overline{\nt}}    
\newcommand{\bpp}{\overline{\pr}}   
\newcommand{\bpI}{\overline{\pr}'}  
\newcommand{\bnI}{\overline{\nt}'}  

\newcommand{\cano}{{{p_0 h_0}}}  
\newcommand{\ope}{{\cal N}}  

\newcommand{\cau}{{\cal U}} 
\newcommand{\caw}{{\cal W}} 
 
\newcommand{\ide}{{\mathbb I}} 
 
\newcommand{\br}{{\bf r}}

\newcommand{\var}{{\cal V} } 
\newcommand{\carI}{$^{12}$C~} 
 
\newcommand{\oxyI}{$^{16}$O~} 
\newcommand{\oxyII}{$^{22}$O~} 
\newcommand{\oxyIII}{$^{24}$O~} 
\newcommand{\caI}{$^{40}$Ca~} 
\newcommand{\caII}{$^{48}$Ca~}

\newcommand{\nicI}{$^{56}$Ni~} 
\newcommand{\nicII}{$^{68}$Ni~}

\newcommand{\tp}{$T_+$\,\,} 
\newcommand{\tm}{$T_-$\,\,} 
 
\newcommand{\hfp}{{\cal Q}}

\newcommand{\tup}{\tau_+} 
\newcommand{\tdo}{\tau_-} 
\usepackage[usenames]{color} 
\usepackage{ulem}

\begin{document} 
 
\noindent 
\title{Self-consistent continuum random-phase approximation with finite-range 
interactions for charge-exchange excitations 
}

\author{V. De Donno} 
\affiliation{Dipartimento di Matematica e Fisica ``E. De Giorgi'', 
  Universit\`a del Salento, I-73100 Lecce, ITALY} 
\author{G. Co'} 
\affiliation{Dipartimento di Matematica e Fisica ``E. De Giorgi'', 
  Universit\`a del Salento, I-73100 Lecce, ITALY and, 
 INFN Sezione di Lecce, Via Arnesano, I-73100 Lecce, ITALY} 
\author{M. Anguiano, A. M. Lallena} 
\affiliation{Departamento de F\'\i sica At\'omica, Molecular y 
  Nuclear, Universidad de Granada, E-18071 Granada, SPAIN} 
\date{\today}

\bigskip 
 
\begin{abstract} 
The formalism of the continuum random-phase approximation theory which
treats, without approximations, the continuum part of the single-particle spectrum, is extended to describe charge-exchange
excitations. Our approach is self-consistent, meaning that we use a
unique, finite-range, interaction in the Hartree-Fock calculations
which generate the single-particle basis and in the continuum random-phase approximation which describes the excitation.  We study
excitations induced by the Fermi, Gamow-Teller and spin-dipole
operators in doubly magic nuclei by using four Gogny-like finite-range
interactions, two of them containing tensor forces. We focus our
attention on the importance of the correct treatment of the continuum
configuration space and on the effects of the tensor terms of the
force.
\end{abstract} 
 
\bigskip 
\bigskip 
\bigskip 
 

\maketitle 

\section{Introduction} 
\label{sec:intro} 
The understanding of  astrophysical nucleosynthesis, especially
that induced by r processes, requires knowledge of charge-exchange
excitations in unstable nuclei which, at least at the moment,
have not been experimentally investigated \cite{arn07}.  For this
reason, there is a remarkable effort to construct nuclear structure
models that can be applied in all the regions of nuclear chart, even
to those nuclei where the experimental information is absent.
 
The random-phase approximation (RPA) theory \cite{boh52}, is one of
the approaches more often used to describe nuclear collective
excitations, and the original formulation for its application to
nuclear systems \cite{row70} has been extended to study
charge-exchange excitations \cite{hal67,lan80,aue81,aue83}.  The
largest part of the first applications of the RPA used
phenomenological inputs. In the spirit of the Landau-Migdal theory of
finite Fermi systems \cite{mig67}, the parameters of the mean-field
potential, generating single-particle (s.p.) wave functions and
energies, and those characterizing the interaction were suitably
chosen to reproduce some experimentally known properties of the
nucleus under study. Despite its success \cite{ost92,har01,ich06}, the
application of this phenomenological approach is limited to those
nuclei whose ground state properties are experimentally known.
 
The so-called self-consistent approaches overtake these limitations.
The s.p. wave functions and energies are produced by Hartree-Fock (HF)
calculations which use the same effective interaction employed in the
RPA. The only input here is the effective interaction, and the values
of the parameters characterizing the force are chosen once forever in
a global fit of ground state properties involving a large set of
nuclei distributed in the various regions of the nuclear chart
\cite{cha07t}. Once the force parameters have been
selected by the fit procedure, the self-consistent calculations are
parameter free and they can be applied to evaluate, and predict, the
properties of every nucleus, independently of the experimental
information about it.
 
Self-consistent studies of charge-exchange excitations have been
conducted mainly with zero-range interactions of Skyrme type
\cite{ham00,fra07}. Recently, these interactions have been implemented
with tensor terms, and the effects of these new terms on
charge-exchange excitations have been studied \cite{bai09a,bai09b,
  bai10, bai11a,bai11b, min13}.  Zero-range effective interactions
have the great merit of simplifying the calculations. There are,
however, various drawbacks in their use, many of them discussed already
in Ref. \cite{dec80} where the D1 parametrization of the finite-range
Gogny interaction was proposed.
 
From the physics point of view, the present article is the natural
continuation of the work of Ref.  \cite{don14b} where we presented
results of charge-exchange responses calculated within the HF plus RPA
(HF+RPA) framework with finite-range interactions.  In that work, the
s.p. configuration space was artificially discretized by setting
boundary conditions at the edge of the radial integration box in the
HF calculation.  The RPA calculations were carried out by using this
discrete s.p. configuration space whose dimensions are large enough
such as the results obtained are independent on the eventual
enlargement of this space. We call this type of
calculations discrete RPA (DRPA) .
 
In Ref. \cite{don14b} we applied our DRPA model to study
charge-exchange excitations in $^{48}$Ca, $^{90}$Zr, and $^{208}$Pb
nuclei, and we obtained satisfactory results in the description of the
experimental centroid energies of these excitations. Since part of the
charge-exchange strength lies above the particle emission threshold,
we wonder whether the use of a discrete configuration space
could affect our results. This worry is more relevant for neutron-rich
nuclei where the particle emission threshold is lower than in the
doubly closed shell nuclei we had investigated.
 
From the theoretical point of view the present work is the natural
extension of Ref. \cite{don11a}, where we introduced an RPA formalism
that fully considers the continuum s.p. configuration space in the
description of charge-conserving excitations. In the present article,
for the reasons discussed above, we extend this continuum RPA (CRPA)
method to the description of charge--exchange excitation modes. To the
best of our knowledge, these are the first self-consistent CRPA
calculations for charge-exchange excitations carried out with
finite-range interactions.

In Sec. \ref{sec:model} we present the set of equations which defines
our HF+CRPA model for charge-exchange excitations. The technical
details of the calculations and the effective interactions used in our
investigation are described in Sec. \ref{sec:details}.  In
Sec. \ref{sec:results} we show, and discuss, the results we have
obtained by applying our model to a set of medium-heavy nuclei, and we
address particular attention to the role played by the tensor force.
Finally, in Sec. \ref{sec:summary} we summarize the main results of
our investigation and we draw our conclusions.
 
\section{The model} 
\label{sec:model} 

The CRPA formalism presented in Refs.
\cite{don08t,don09,don11a,don11b,co13} is constructed to handle the
continuum s.p. configuration space without approximation, even when
finite-range interactions are used.  Its extension to charge-exchange
excitations does not require new hypotheses, or approximations;
however, it is not straightforward and it deserves to be presented with
some detail.

The basic idea of our formalism is to rewrite the usual RPA secular
equations \cite{rin80}, where the unknown variables are the $X$ and
$Y$ amplitudes, in terms of new unknowns called channel functions
defined as
\beq  
f^\cano_{ph}(r) \, = \, \sumint_{\epsilon_p} \,  
X^\cano_{ph}(\epsilon_p) \, R_p(r,\epsilon_p) \, =\, 
R_{p_0}(r,\epsilon_{p_0})\, \delta_{p p_0}\, \delta_{hh_0} \, 
+ \, \sum_\mu \, c^{\mu+}_{ph} \, \widetilde{\Phi}^{\mu+}_{p}(r) 
\label{eq:ff} 
\eeq  
and 
\beq  
g^\cano_{ph}(r) \, = \, \sumint_{\epsilon_p} \,  
\left[Y^{\cano}_{ph}(\epsilon_p)\right]^* \, R_p(r,\epsilon_p) \,=\, 
\sum_\mu \, c^{\mu-}_{ph} \, \widetilde{\Phi}^{\mu-}_{p}(r) \, .
\label{eq:gg} 
\eeq  
The definition of the $f$ and $g$ functions is given by the first
equalities where the symbol $\sum \!\!\!\!\!\!\displaystyle\int $
indicates the sum on discrete s.p. energies and the integration on the
continuum part of the spectrum, and $R_p$ is the radial part of the
particle wave function.  The label $p_0h_0$ indicates the {\sl elastic
  channel}, defined as the specific channel where the particle is
emitted.  The number of elastic channels does not, usually, coincide
with that of the particle-hole ($ph$) pairs, since the particle
channel must be open; in other words, in an elastic channel the energy
of the particle state must be positive, i.e. in the continuum.  This
implies that the excitation energy $\omega$ of the full system must be
larger, in absolute value, than the energy of the hole state
$\epsilon_h$.  The CRPA equations are solved by imposing, every time,
that the particle is emitted in a different elastic channel.

The second equalities of Eqs. \eqref{eq:ff} and \eqref{eq:gg} indicate
that, in our approach, we expand the channel function on a basis.
Specifically, we use a set of orthonormal Sturm functions \cite{raw82}
\beq
\widetilde{\Phi}^\mu_{p}(r) \,= \, 
\Phi^\mu_{p}(r)\, - \, \sum_{\epsilon_i<\epsilon_{\rm F}} \, \delta_{l_i,l_p} \,\delta_{j_i,j_p} \,
R^*_i(r) \, \int {\rm d}r' \, r'^2 \, R_i(r') \, \Phi^\mu_p(r')
\, 
\label{eq:orthost}
\eeq 
In Eqs. \eqref{eq:ff} and \eqref{eq:gg} the superscripts $+$ and $-$
indicate that the Sturm functions are calculated for
$\epsilon_p=\epsilon_h+\omega$ or $\epsilon_p=\epsilon_h-\omega$,
respectively, and we have dropped the explicit dependence on the open
channel label $p_0h_0$ of all the $c^\mu_{ph}$ expansion coefficients
to simplify the writing.
 
The charge-exchange excitations can be classified as isospin lowering
$T_-$, when the hole is a neutron and the particle is a proton, and
isospin rising $T_+$, in the opposite case.  We use the convention of
indicating with $\pr$ and $\nt$ a proton and a neutron state,
respectively, and with a bar a hole state. Therefore, we have $\pr
\bn$ pairs in \tm, and $\nt \bpp$ pairs in \tp excitations.

We show in the Appendix \ref{sec:appmat} that charge-exchange CRPA
($pn$-CRPA) secular equations for \tm excitations can be expressed as
\beqn 
\begin{bmatrix} 
A^{\mu +}_{\pi \bn,\pi' \bn'} & -B^{\mu +}_{\pi\bn,\nu' \bpp'} \\  
-(B^{\mu -}_{\nu\bpp,\pi' \bn'})^* & (A^{\mu -}_{\nu\bpp,\nu' \bpp'})^*  \\  
\end{bmatrix} 
\begin{bmatrix} 
c^{\mu +}_{\pi' \bn'}\\  
(c^{\mu -}_{\nu' \bpp'})^* \\ 
\end{bmatrix}= 
\begin{bmatrix} 
C_{\pi \bn,\pi_0\bn_0}\\ 
(D_{\nu \bpp,\pi_0\bn_0})^*\\ 
\end{bmatrix}  
\label{eq:pnCRPA1} 
\, , 
\eeqn 
and for \tp excitations as
\beqn 
\begin{bmatrix} 
A^{\mu +}_{\nu\bpp,\nu' \bpp'} & -B^{\mu +}_{\nu\bpp,\pi' \bn'}  \\  
-(B^{\mu -}_{\pi\bn,\nu' \bpp'})^* & (A^{\mu -}_{\pi\bn,\pi' \bn'})^* \\  
\end{bmatrix} 
\begin{bmatrix} 
c^{\mu +}_{\nu' \bpp'}\\  
(c^{\mu -}_{\pi' \bn'})^* \\ 
\end{bmatrix}= 
\begin{bmatrix} 
C_{\nu \bpp,\nu_0\bpp_0}\\ 
(D_{\pi \bn,\nu_0\bpp_0})^*\\ 
\end{bmatrix}
\,.
\label{eq:pnCRPA2} 
\eeqn 

The explicit expressions of the $A$ and $B$ interaction matrix
elements of the above equations are given in Appendix
\ref{sec:appmat}, where it is shown that they depend on the excitation
energy. Since Eqs. (\ref{eq:pnCRPA1}) and (\ref{eq:pnCRPA2}) are,
separately, inhomogeneous sets of linear algebraic equations, they
have solutions, different from the trivial one, for each value of
$\omega$ above the nucleon emission threshold.  As already stated,
these equations are solved for each elastic channel, whose number is
always equal to that of the open channels.  The knowledge of the
expansion coefficients $c^{\mu \pm}$ allows us to reconstruct the
channel functions, as indicated by Eqs. \eqref{eq:ff} and
\eqref{eq:gg}.  In particular, for a \tm excitation we have $c^{\mu
  +}_{\nu' \bpp'}=c^{\mu -}_{\pi' \bn'}=0$, therefore only $f_{\pr'
  \bnI}^{\pi_0 \bn_0}$ and $g_{\nt' \bpI}^{\pi_0 \bn_0}$ are different
from zero.  In contrast, if the reaction is of \tp type, we have
$c^{\mu +}_{\pi' \bn'}=c^{\mu -}_{\nu' \bpp'}=0$, consequently only
$f_{\nt' \bpI}^{\nu_0 \bpp_0}$ and $g_{\pr' \bnI}^{\nu_0 \bpp_0}$ are
different from zero.
 
For a given excitation energy, and for each elastic channel, the full
set of CRPA function channels $f$ and $g$ allows us to calculate the
nuclear response induced by an external operator $\hfp^{\ope
  \pm}_{J^\Pi,M}$.  In our calculations, we consider one-body
operators of the form
\beq 
\hfp^{\ope \, \pm}_{J^\Pi,M}(\br)\, = \, \sum_{i=1}^A \, \kappa^\ope_J(r_i) 
\, \theta^\ope_{J^\Pi,M}(\Omega_i) \, \delta(\br_i-\br)\, t_{\pm}(i)\, ,  
\eeq 
where the dependence on the radial, angular-spin, and isospin quantum
numbers is factorized.  For the transition matrix element we obtain
the expression
\beqn 
\Gamma_{J^\Pi}^{\ope\,\pm}(\omega) &=&   
\, \sum_{p_0h_0} \Big|\sum_{p h}\,   
\left[ \langle j_p \| \theta^\ope_{J^\Pi} \| j_h \rangle \, 
\langle \frac{1}{2} t_p | t_{\pm} | \frac{1}{2} t_h \rangle \,   
\int {\rm d}r \, r^2 \, (f^\cano_{ph}(r))^* \, \kappa^\ope_J(r) \, R_h(r)   
\right.
\\ 
&~& 
\nonumber
\left. + \, (-1)^{J+j_p -j_h} \,  
\langle j_h \| \theta^\ope_{J^\Pi} \| j_p \rangle \,   
\langle \frac{1}{2} t_h | t_{\pm} | \frac{1}{2} t_p \rangle \, 
\int {\rm d}r \, r^2 \,  R^*_h(r) \, \kappa^\ope_J(r) 
\, g^\cano_{ph}(r) \right]\Big|^2 
\, , 
\label{eq:transs1} 
\eeqn  
where with the double bar we indicate the reduced matrix elements, as
defined in \cite{edm57}.  In the previous equations we used $t_\pm =
\tau_\pm /2$, where $\tup$ and $\tdo$ are the isospin operators
transforming, in our convention, a proton into a neutron and vice
versa.
 
In the present work we consider the Fermi ($\ope \equiv {\rm F}$), 
\beq 
\hfp_{0^+,0}^{{\rm F}\pm}\, = \, \sum_{i=1}^A \,t_{\pm}(i) 
\label{eq:f} 
\, , 
\eeq 
and the Gamow-Teller ($\ope \equiv {\rm GT}$),  
\beq 
\hfp_{1^+,M}^{{\rm GT}\pm} \,= \, \sum_{i=1}^A  \, \bsigma_M(i) \, t_{\pm}(i)\, = \,  
\sqrt{4\pi} \, \sum_{i=1}^A  \, [Y_0(i) \otimes \bsigma(i) ]^1_M \, t_{\pm}(i) 
\, ,
\label{eq:gt} 
\eeq 
operators that excite $0^+$ and $1^+$ states, respectively. In the
above equation we have indicated with $Y_{L}$ the spherical harmonics
and with $\bsigma$ the Pauli matrix operator acting on the spin
variable. The symbol $\otimes$ indicates the usual tensor product
between irreducible spherical tensors \cite{edm57}.

Finally, we have also studied the excitations induced by the spin
dipole ($\ope \equiv {\rm SD}$) operator 
\beq \hfp_{J^-,M}^{{\rm
    SD}\pm} \, = \, \sum_{i=1}^A \, r_i \, [Y_1(i)\otimes \bsigma(i)
]^J_M \, t_{\pm}(i)
\label{eq:sd} 
\, , \eeq 
which excites the multipoles $0^-$, $1^-$ and $2^-$. In this
case, we calculate the strength functions corresponding to each
individual multipolarity and also the total SD strength
\beq 
\Gamma^{{\rm SD}\pm}(\omega) \, = \, \sum_{J^\Pi=0^-,1^-,2^-} 
\Gamma_{J^\Pi}^{{\rm SD}\pm}(\omega)  
\,\,. 
\label{eq:sumSD} 
\eeq 

In our study we have verified the sum-rule exhaustion, an important
tool to investigate the global properties of the charge-exchange
excitations. For this purpose, we have defined the sum-rule exhaustion
function
\beq
{\rm SR}_{J^\Pi}^{\ope\,\pm}(\omega)\, =\, \int_0^\omega {\rm d}\omega' 
\, \Gamma_{J^\Pi}^{\ope\,\pm}(\omega') \, .
\label{eq:exha}
\eeq

We also found convenient to formulate the sum rules in terms of energy
moments defined as
\beq 
m_{\lambda}^{\ope\,\pm}\, =
\, \sum_{J^\Pi} \, m_{\lambda}(\Gamma^{\ope \, \pm}_{J^\Pi}) \, , 
\label{eq:mom} 
\eeq 
with 
\beq 
m_{\lambda}(\Gamma^{\ope\,\pm}_{J^\Pi})\, = \, 
\int_0^\infty {\rm d}\omega \, \omega^\lambda 
\, \Gamma_{J^\Pi}^{\ope \, \pm}(\omega)  
\, .
\label{eq:momJ} 
\eeq 
According to these expressions, we define the centroid energy of an
excitation induced by the $\ope$-type operator as
\beq 
{\omega}_{\rm cen}^{\ope \, \pm} \, =
\, \frac{m_{1}^{\ope \, \pm}}{m_{0}^{\ope \, \pm}} \,   ,  
\label{eq:centr} 
\eeq 
and the corresponding variance as
\beq 
\var^{\ope \, \pm} \, = \, 
\frac{m_{2}^{\ope \, \pm}}{m_{0}^{\ope \, \pm}} \, -\, 
\left( \frac{m_{1}^{\ope \, \pm}}{m_{0}^{\ope \, \pm}} \right)^2 \, . 
\label{eq:var}  
\eeq 
In the case of the SD transitions, we have calculated the centroid of
the distributions of the individual multipolarities,
\beq 
{\omega}_{{\rm cen},J^\Pi}^{{\rm SD}\pm} \,  
=\, \frac{m_{1}(\Gamma^{{\rm SD}\pm}_{J^\Pi})}{m_{0}(\Gamma^{{\rm SD}\pm}_{J^\Pi})} \,   ,  
\label{eq:centr-mul} 
\eeq 
and also their corresponding variances.  We carried out the numerical
evaluations of Eq. (\ref{eq:momJ}) in a restricted energy range of
which we shall indicate the minimum $(E_<)$, and the maximum $(E_>)$
values.
  
 
\section{Details of the calculations} 
\label{sec:details} 
The only input required by our self-consistent approach is the
effective nucleon-nucleon force.  In our calculations we consider a
two-body nucleon-nucleon (NN) interaction of the form
\beq 
V(\br_i,\br_j) \,=\, \sum_{\eta} v_\eta(|\br_i-\br_j|)\, O^\eta_{i,j}  
\, , 
\label{eq:force1}  
\eeq 
where $v_\eta$ are scalar functions of the distance between the two
interacting nucleons, and $O^\eta$ indicates the type of operator
dependence:
\beqn 
O^{\eta}_{i,j}&:& 1\,,\,\,\btau(i)\cdot\btau(j)\,,\,\, 
                 \bsigma(i)\cdot\bsigma(j)\,,\,\, 
 \bsigma(i)\cdot\bsigma(j)\,\btau(i)\cdot\btau(j)\,, \nonumber 
\\ &~& 
S(i,j)\,,\,\, S(i,j) \btau(i)\cdot\btau(j) 
\,,\,\, {\bf l}_{ij} \cdot {\bf s}_{ij}   
\,,\,\, {\bf l}_{ij} \cdot {\bf s}_{ij}\, \btau(i)\cdot\btau(j)   
\label{eq:fcahnnels} 
\, . 
\eeqn  
In the above expression we have indicated with $\btau$ the Pauli
matrix operator acting on the isospin variable, with ${\bf s}_{ij}$
the total spin of the interacting nuclear pair, with ${\bf l}_{ij}$
its orbital angular momentum, and with
\beq 
S(i,j)\, = \,3 \, \frac {[\bsigma(i)\cdot (\br_i-\br_j)] \, 
                    [\bsigma(j)\cdot (\br_i-\br_j)] }  
                 {(\br_i-\br_j)^2}\, 
- \, \bsigma(i)\cdot\bsigma(j) 
\eeq 
the usual tensor operator. The terms $\eta=7,8$ include the spin-orbit
contributions of the force.
 
In this work we carried out calculations with four Gogny-like
interactions: the D1M force \cite{gor09}, the more traditional D1S
\cite{ber91} parametrization, and also with other two forces, D1MT2c
and D1ST2c \cite{don14b}, which we built by adding tensor terms to the
two basic parametrizations.
 
For complete self-consistent calculations, also the Coulomb and
spin-orbit terms of the interaction should be considered in both steps
of our approach, i.e., the HF and the CRPA calculations.  The Coulomb term
is obviously not active in charge-exchange excitations.  The
spin-orbit term is neglected in our CRPA calculations. We have
recently studied the relevance of these two terms of the interaction
in charge-conserving HF+DRPA calculations and we found that their
contribution is very small \cite{don14a}.
  
The first step of our calculations consists of constructing the
s.p. basis by solving the HF equations with the bound-state boundary
conditions at the edge of the discretization box.  The technical
details concerning the iterative procedure used to solve the HF
equations for a density-dependent finite-range interaction can be
found in Refs. \cite{co98b,bau99}.  In our HF+CRPA formalism, we solve
the HF equations only for those states under the Fermi surface.
 
The second step is the solution of the CRPA equations. The formalism
developed in the previous section leads to a set of algebraic
equations whose unknowns are the expansion coefficients
$c^{\pm}_{ph}$. The number of coefficients, and therefore the
dimensions of the complex matrix to diagonalize, is a numerical input
of our approach.  Since the expansion on a basis of Sturm functions is
a technical artifact, the solution of the CRPA secular equations must
be independent of the number of expansion coefficients. We tested the
convergence of our results by controlling the values of the total
photoabsorption cross section in \oxyI and \caI nuclei. We reached
stability up to the fifth significant figure with ten expansion
coefficients, independently of the multipolarity and of the energy of
the excitation \cite{don09}.
 
With this HF+CRPA model, we have carried out calculations for the
$^{12}$C, $^{16}$O, $^{22}$O, $^{24}$O, $^{40}$Ca, $^{48}$Ca,
$^{56}$Ni, and $^{68}$Ni.  In these nuclei the hole s.p. levels are
fully occupied and this fact eliminates deformations and minimizes
pairing effects.
 
\begin{table}[!b] 
\begin{center} 
\begin{tabular}{c  cccc c cccc}  
\hline \hline 
& \multicolumn{4}{c} {proton} &~~~~~~& \multicolumn{4}{c} {neutron} \\ \cline{2-5} \cline{7-10}
nucleus                & s.p. level         & D1M   & D1Mt2c & exp& & s.p. level  &  D1M  & D1Mt2c & exp\\ \hline
$^{12}$C & $1p_{3/2}$ & 13.82 &   13.56 & 15.96  && $1p_{3/2}$   &  16.36 &  16.09 &   18.72\\ 
$^{16}$O & $1p_{1/2}$ & 11.94 &  12.66  & 12.13  && $1p_{1/2}$   &  15.14 &  15.88 &   15.66\\ 
$^{22}$O & $1p_{1/2}$ & 23.67 &  25.49  &  23.26 && $1d_{5/2}$   &   6.38 &   6.80  &   6.67\\ 
$^{24}$O & $1p_{1/2}$ & 25.69 &  27.65  &  28.64 && $2s_{1/2}$   &  4.11 &  4.07 &  4.19\\ 
$^{40}$Ca & $1d_{3/2}$ & 8.86 &  9.58  &   8.32  && $1d_{3/2}$  &  15.74 &  16.49 &  15.65\\ 
$^{48}$Ca & $1d_{3/2}$ & 16.69 & 18.38  &  15.81 && $1f_{7/2}$   &  9.33 &   9.72 &  9.94\\ 
$^{56}$Ni  & $1f_{7/2}$ & 6.86 &   6.54     &   7.16  && $1f_{7/2}$   &  16.00 &  15.67 &  16.62 \\ 
\hline \hline 
\end{tabular}
\vspace*{-2mm}
\caption{Energy thresholds, in MeV, for proton and neutron emission 
of the nuclei considered in this work. The experimental values have been 
obtained as explained in the text by using the binding energies values 
taken from the compilations of Refs.  \cite{bnlw,aud03}.
} 
\label{tab:the} 
\end{center} 
\end{table}  

\section{Results} 
\label{sec:results} 

In this section we present some results obtained in our investigation
of charge-exchange excitations induced by the F, GT, and SD operators
in the aforementioned nuclei. We shall be concerned about the part of
the strength lying above the emission particle threshold.

We give in Table \ref{tab:the} the energy thresholds for proton and
neutron emission, for the various nuclei we have considered in the
present work. In our model, these values are those of the
s.p. energies of the last occupied level. By using the physical
interpretation of s.p. energies given by Koopman's theorem
\cite{rin80}, we have calculated the experimental values as differences
between the binding energies of the indicated nucleus and those of the
nuclei with one fewer nucleon. The binding energies have been taken from
the compilations of Refs.  \cite{bnlw,aud03}.  This comparison with
the experimental values gives an idea of the level of accuracy of our
mean-field model in the description of measured quantities.  The
quality of our approach in the description of the ground state
properties of the nuclei under investigation is discussed in detail in
Ref. \cite{co12b}.

Before discussing the RPA results, we point out some features of our
approach related to the need of the proper treatment of the continuum
part of the s.p. configuration space.  All the calculations have been
carried out by using the four interactions presented in
Sec. \ref{sec:details}, but, to simplify the discussion, we present
here only the results obtained with the D1M and D1MT2c forces, since
the results obtained with the other two forces do not show relevant
differences.

\begin{figure}[!t] 
\begin{center} 
\includegraphics [scale=0.5,angle=0]{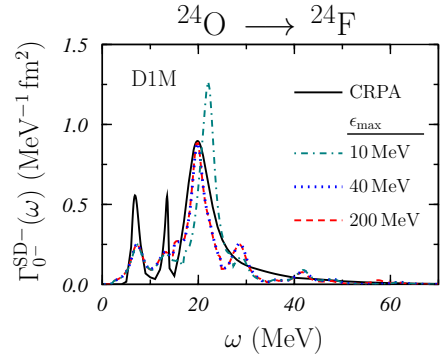} 
\vspace*{-0.3cm}
\caption{\small The contribution of the 0$^-$ multipole to the \tm SD
  response in $^{24}$O, calculated in DRPA for different values
  of the s.p cut off energy $\epsilon_{\rm max}$ compared with 
  the CRPA result shown by the full black line. The
  DRPA results are folded with a Lorentz function of 3 MeV width.
}
\label{fig:D1M_conv} 
\end{center} 
\end{figure} 
\begin{figure}[!b] 
\begin{center} 
\includegraphics [scale=0.5,angle=0]{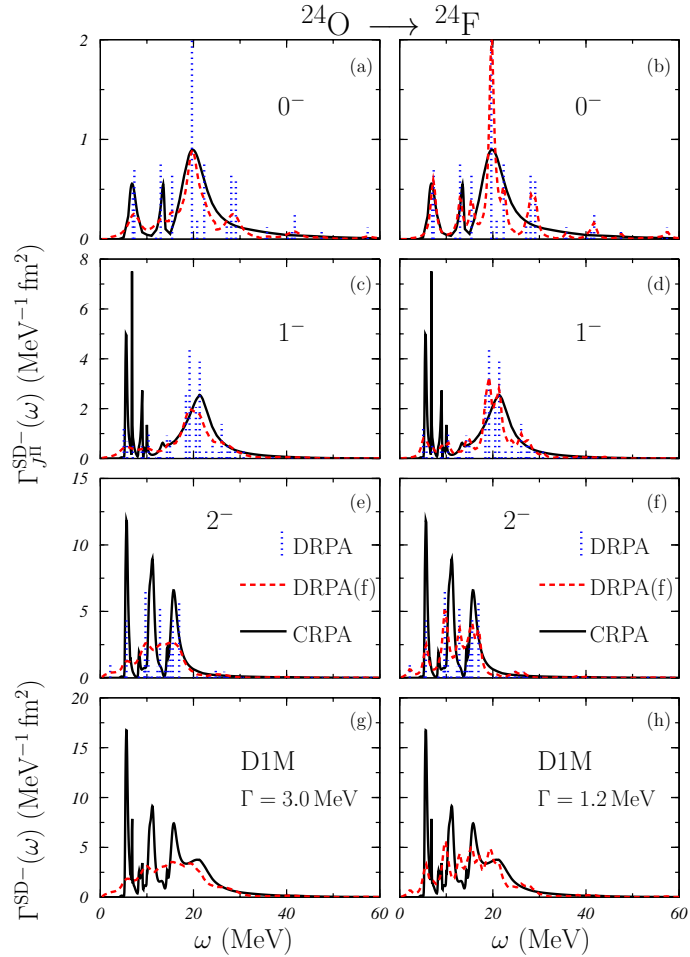} 
\vspace*{-0.3cm}
\caption 
{\small Energy distributions of the \tm SD strengths for the \oxyIII
  nucleus obtained with the D1M interaction.  In the upper panels 
  we show, separately, the strengths of the $0^-$, $1^-$ and
  $2^-$ excitations, and in the two lowest panels, (g) and (h),
  their sums.
  The CRPA results are represented by the black solid curves.  The
  vertical dotted blue lines show the DRPA results that generate the 
  dashed red lines after a folding with a Lorentz function.  The results
  obtained by using, in the folding procedure, a 3 MeV width are shown 
  in the left panels, i.e. the (a), (c), (e) and (g) panels, and the results
  obtained with 1.2 MeV width are shown in the right panels. 
}
\label{fig:disc} 
\end{center} 
\end{figure} 

\newpage
\phantom{blablablabla}

We show in Fig. \ref{fig:D1M_conv} the strength function for the \tm
excitation of the $0^-$ state induced by the SD operator in
$^{24}$O. This is a typical situation we have encountered. The full
line represents the CRPA results obtained with the D1M interaction,
and all the other lines the DRPA results folded with a Lorentz
function of 3 MeV width.  The various lines show the sensitivity of
the DRPA results to the choice of the maximum energy of the
s.p. configuration space, $\epsilon_{\rm max}$. The figure clearly
shows that the convergence of the DRPA results is reached already for
$\epsilon_{\rm max}$ = 40 MeV. The folding of the DRPA results cannot
reproduce the smooth behavior of the CRPA strength above 20 MeV.

The limitations of the folding procedure are clarified in
Fig. \ref{fig:disc} where we show the results related to all the
multipole excitations induced by the SD operator in the \oxyIII
nucleus. In this figure, we indicate the original DRPA results, at
convergence, with dotted vertical lines, to be compared with the full
black curves representing the CRPA results.  For each multipolarity,
the position of the DRPA main peaks coincides with the maxima of the
CRPA results. The two descriptions are rather similar in the lower
energy region, but above 20 MeV, the DRPA produces peaks, while the
CRPA responses have a rather smooth behavior.  These peaks are
clearly a consequence of the use of discrete s.p. basis.

The dashed red lines of the figure have been obtained by folding the
DRPA with Lorentz functions.  In the left panels the results have been
obtained by using a width of 3 MeV, chosen to reproduce at best the
CRPA $0^-$ response, while in the right panels the width is of 1.2 MeV
because this value is more adequate to reproduce the CRPA $2^-$
response.  The first choice generates a too large smoothing for the
$2^-$ state, which is the dominant contribution, therefore the total
SD$^-$ response is much more spread than that of the CRPA. The second
choice produces a global strength more similar to that of the CRPA,
but fails in describing the CRPA results of the $0^-$ and $1^-$
excitations.

After having clarified the relation between the DRPA and CRPA results,
we shall discuss, in order, the GT, F and SD excitations
of the nuclei under investigation.  In an extreme independent particle
model (IPM), $N=Z$ nuclei with all the spin-orbit partner levels
occupied cannot produce GT excitations. In our approach, which
considers 1p-1h excitations only, the GT excitations for these nuclei
are not prohibited but strongly hindered. The GT strength measured in
$^{16}$O \cite{mer94} is about one order of magnitude larger than our
RPA strength. As pointed out in Ref. \cite{bro94} only calculations
considering, at least, 4p-4h excitations can produce the correct
amount of strength.

\begin{figure}[!b] 
\begin{center} 
\includegraphics [scale=0.6,angle=0]{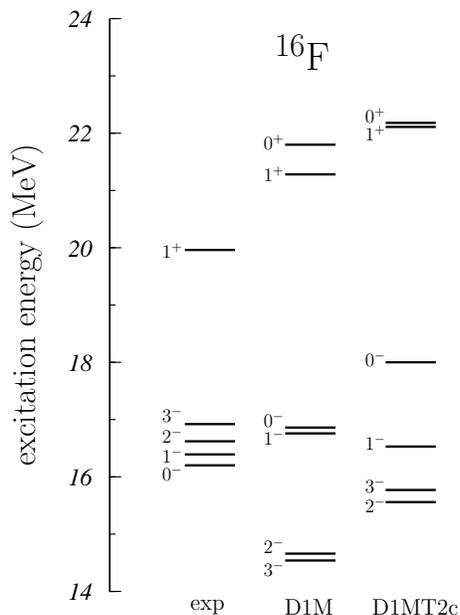} 
\vspace*{-0.3cm}
\caption 
{\small Spectrum of the $^{16}$F nucleus as obtained from charge-exchange
experimental experiments \cite{faz82,mer94,mad97,fuj09}. 
The energies are relative to the \oxyI ground state.
}
\label{fig:spectrum} 
\end{center} 
\end{figure} 

In Fig. \ref{fig:spectrum} we compare the results of our RPA
  calculations with the spectrum of $^{16}$F obtained in
  charge-exchange experiments on the \oxyI target
  \cite{faz82,mer94,mad97,fuj09}. The order of magnitude of the
  excitation energies is reproduced, but the detailed structure of
  this spectrum is not well described. This requires the inclusion of
  excitations beyond 1p-1h.

\begin{figure}[!t] 
\begin{center} 
\includegraphics [scale=0.5]{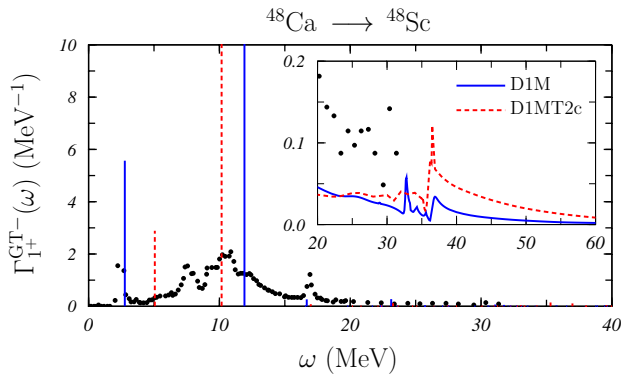} 
\vspace*{-0.3cm}
\caption{\small  
Energy distributions of the \tm GT strength for the \caII
nucleus. The vertical lines show the results of the DRPA
calculations carried out with the D1M (full blue lines) and 
D1MT2c (dashed red lines)
interactions, respectively. 
In the inset we show the CRPA results. The full blue and dashed red
curves indicate the CRPA results for the D1M and D1MT2c interactions,
respectively.
The experimental data are taken from Ref. \cite{yak09}.
}
\label{fig:GTca48} 
\end{center} 
\end{figure}

Our RPA calculations are more adequate for nuclei where the GT
excitation is related mainly to a 1p-1h excitation. We show in Fig.
\ref{fig:GTca48} the energy distribution of the \tm GT for $^{48}$Ca.
The experimental data are those of Ref. \cite{yak09}.  The full blue and
dashed vertical lines show the DRPA results for the D1M and D1MT2c
forces, respectively.

The GT responses are dominated by the peaks shown in the main part of
the figure. The strengths in the region above these main peaks are
much smaller. We present in the inset of
Fig. \ref{fig:GTca48}, the responses of the CRPA calculation
in the region above 20 MeV, by using a different scale. 
The CRPA results obtained with D1M and D1MT2c
interactions are indicated by the full blue and dashed red curves,
respectively.

The DRPA calculations exhaust the so-called Ikeda sum rule at 1\%
level for the D1MT2c force and much better, to one part in over 100000,
when the D1M interaction is used.  For each of the two
interactions, the DRPA calculations generate two peaks. The peak
located at the lower energy, below the emission particle threshold, is
dominated by the transition from the neutron $1f_{7/2}$ level to the
bound proton $1f_{7/2}$ level.  The second peak, whose excitation
energy is above the emission particle threshold, is dominated by 
the transition from the neutron $1f_{7/2}$ level to the proton
$1f_{5/2}$ level, which, in our calculations, is bound.  Such
RPA solutions with energy eigenvalues in the continuum but dominated
by bound s.p. transitions are not well described by our CRPA
formalism, tailored for the continuum.  A meaningful comparison
between DRPA and CRPA results should be done in the region beyond the
two main peaks.  In the region above 20 MeV, the DRPA and CRPA
total strengths are very similar.  The
global DRPA strength of the \tm GT transition for
the D1M interaction is 24.17. The contribution of the region above
the main peak is 1.05 to be compared with the CRPA 
integrated strength of 1.08.  The 
agreement between DRPA and CRPA results for the D1MT2c
interaction is not so precise.  The total DRPA \tm GT strength is
24.04 of which  1.18 is beyond the main
peak. In this region, the CRPA total strength is 1.58. 

The width of the experimental strength is much larger than that of the
RPA calculations.  We have already encountered this kind of problems
in the comparison of our charge-conserving CRPA results with total
photoabsorption cross sections \cite{don11a}. It is a common feature
of the RPA description of the giant resonances \cite{spe91} and it is
attributed to its intrinsic limitation of considering one-particle
one-hole excitations only.  It is evident that the RPA strength needs
a redistribution which will be provided by the inclusion of 2p-2h
degrees of freedom \cite{gam12}.  The results presented in the inset
show a sudden increase of the CRPA responses above the 35 MeV due to
the opening of the channels related to the emission of neutrons from
the $1s_{1/2}$ s.p. level. This sharp peak of the CRPA response would
be, probably, smoothed by the inclusion of 2p-2h excitations.

The total experimental strength is $15.36 \pm 2.16$, corresponding to
about 60\% of the theoretical strength only. The measurements stop at
an excitation energy of about 30 MeV, and our calculations indicate
that there is GT strength above this energy.  The presence of strength
above the maximum energy measured in the experiment is probably the
explanation of the missing GT strength.  For example, analysis of the
$^{90}$Zr (p,n) $^{90}$Nb data \cite{ich06} indicates that all the GT
strength is present in the excitation, much of it above the main peak.

The difference between the positions of the main peaks of the DRPA
results in $^{48}$Ca, calculated with the D1M and D1MT2c forces is
essentially due to the effect of the tensor force on the energy of the
proton $1f_{5/2}$ level.  Otsuka and collaborators \cite{ots05}
pointed out that the tensor force when acting on s.p. levels with
different isospin lowers the energy of the state with $j=l-1/2$ and
increases that of the state with $j=l+1/2$.  In this case the energy
of the proton $1f_{5/2}$ level is lowered, and that of the $1f_{7/2}$
is enhanced by the tensor force.  The RPA results shown in
Fig. \ref{fig:GTca48} are clearly affected by this effect.  The energy
of the first excited state, where the main p-h transition is on the
proton $1f_{7/2}$ level, is increased, while that of the second state,
that with the $1f_{5/2}$ level, is lowered. The effect of the tensor
force on the RPA calculation in itself is negligible as compared with
the effect on the s.p. energies.

\begin{figure}[ht] 
\begin{center} 
\includegraphics [scale=0.5]{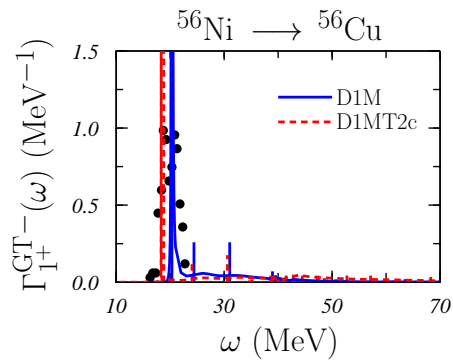} 
\vspace*{-0.3cm}
\caption{\small
Energy distributions of the \tm GT strength for the 
$^{56}$Ni nucleus. The meaning of the lines is the same 
as that of Fig. \ref{fig:GTca48}. The data are taken from 
Ref. \cite{sas11}.
}
\label{fig:GTni56} 
\end{center} 
\end{figure} 

The points discussed for the GT excitation in \caII can also be
considered for the \nicI nucleus. We show in Fig. \ref{fig:GTni56} the
energy distribution of the \tm GT strengths obtained with our DRPA and
CRPA calculations and we compare them with the experimental data of
Ref. \cite{sas11}.  The meaning of the various lines is as in the previous figure.

In contrast to $^{48}$Ca, the \nicI nucleus is a $N=Z$ nucleus, and,
differently from \oxyI and \caI, not all the spin-orbit partner levels
are occupied. For this reason there are large GT excitation strengths
from the $1f_{7/2}$ hole to the $1f_{5/2}$ particle state, in both the
neutron-proton and proton-neutron cases. In effect the zero value
predicted by the Ikeda sum rule for the $N=Z$ nuclei is obtained by
subtracting the \tm and \tp total strengths which are about 12,
much larger than the values $\sim 0.05$ and $\sim 0.1$ we found in
\oxyI and \caI, respectively.

Differently from the \caII case, we observe in Fig. \ref{fig:GTni56}
that also the CRPA strength distributions present large peaks
coinciding with the main peaks of the DRPA results. This is because
the $1f_{5/2}$ s.p. states are unbound for both protons and neutrons,
independently of the interaction, considered.  Our CRPA formalism
treats these states in the continuum, even though they appear as sharp
resonances and produce rather narrow peaks in the energy
distributions.

The comparison with the data indicates the need to consider in the
calculations also the spreading width.  The integrated experimental
strength is about 3.6, i.e roughly the 30\% of our RPA strength. On
the other hand, the results of the figure show that there is
considerable strength above the maximum measured energy.  In our
calculations this strength is about 10\% of the total one in
calculations both with and without tensor.  Finally, we should recall that
in our model \nicI is a doubly closed shell nucleus, but there are
indications that pairing effects could be relevant \cite{bai13}.


In analogy with what we have already discussed for the GT transitions,
also the F transitions are hindered in $N=Z$ nuclei. For nuclei with
neutron excess the main peak is concentrated in the transition between
the occupied neutron s.p. state and the empty isobaric analog proton
state.  Our DRPA calculations identify strong 0$^+$ states at 3.35,
3.28, and 6.96 MeV in \oxyII, \oxyIII and \caII nuclei, respectively.  All
these states are below the continuum energy threshold.  Unfortunately,
in the low-energy spectra of the $^{22}$F, $^{24}$F and $^{48}$Sc
nuclei these states have not been identified. In the nuclei with
neutron excess we have investigated, the \tm F strength allocated in
the continuum region is less than 1\% of the total one.  For the $N=Z$
nuclei studied, all the \tm F strengths appear above the continuum
threshold, and their total values are comparable to those found in the
continuum part of strength of the nuclei with neutron excess.

\begin{figure}[!b] 
\begin{center} 
\includegraphics [scale=0.4,angle=0]{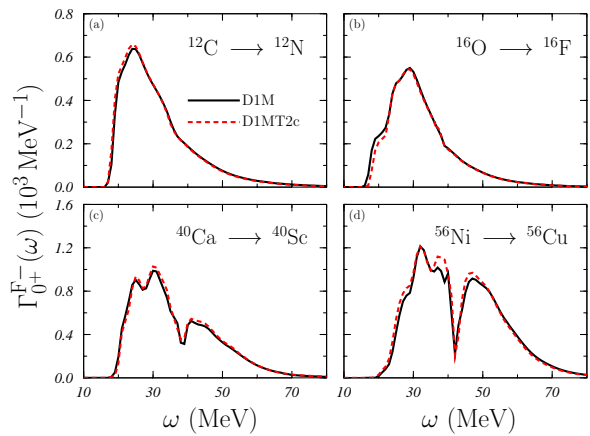} 
\vspace*{-0.3cm}
\caption{\small  
Energy distributions of the \tm F responses obtained in CRPA 
calculations with the  D1M (full black lines) and 
D1MT2c (dashed red lines) interactions.  
}
\label{fig:Fermi} 
\end{center} 
\end{figure} 

As example of our results, we show in Fig. \ref{fig:Fermi} the \tm F
responses for the \carI, \oxyI, \caI and \nicI nuclei which have the
same number of protons and neutrons.  The responses of the $^{40}$Ca
and $^{56}$Ni nuclei show two large peaks. These are due to the
importance of different particle-hole transitions in the RPA
response. In $^{56}$Ni the lower energy peak is characterized by a p-h
pair related to the $f_{7/2}$ s.p. states, while the peak at higher
energy by a p-h pair related to the $d_{3/2}$ hole.  In the case of
the $^{40}$Ca, both peaks are dominated by the $d_{5/2}$ p-h pair. The
residual interaction in the RPA theory mixes this pair with the
$d_{3/2}$ p-h pair in the lower energy peak and with the $p_{3/2}$ p-h
pair in the higher energy peak.
In the figure the full black lines show the F strengths obtained with
the D1M interaction and the dashed red lines those found with D1MT2c.
In this case, the tensor force does not produce remarkable effects.


\begin{figure}[!b] 
\begin{center} 
\includegraphics [scale=0.5,angle=0]{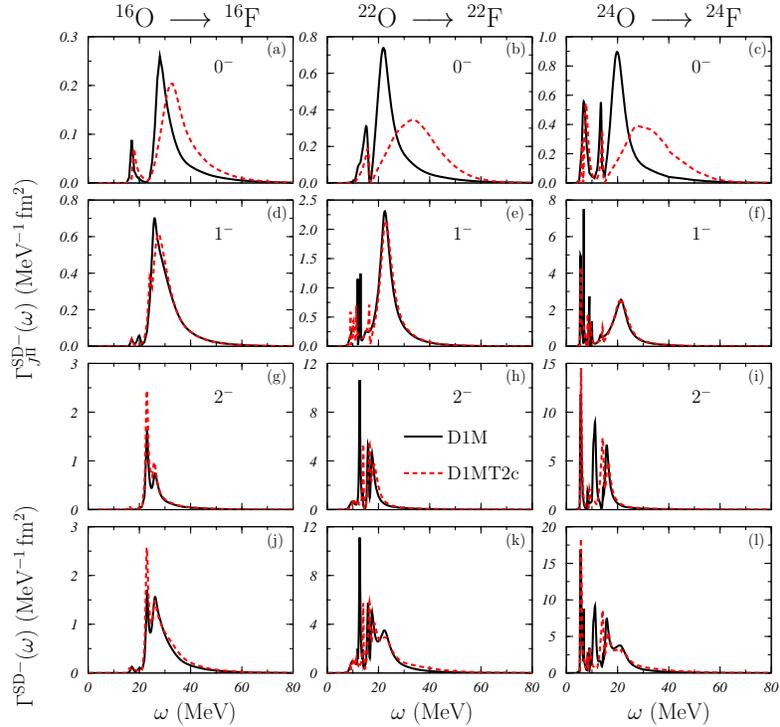} 
\vspace*{-0.3cm}
\caption{\small Energy distributions of the \tm 
$\Gamma_{J^\Pi}^{{\rm SD}}(\omega)$ and  
$\Gamma^{{\rm SD}}(\omega)$ strengths for \oxyI, \oxyII and \oxyIII nuclei 
obtained in CRPA calculations.   
The solid black curves have been obtained with the D1M interaction  
while the dashed red curves with the D1MT2c force. 
} 
\label{fig:SD_oxy} 
\end{center} 
\end{figure} 

\begin{figure}[!t] 
\begin{center} 
\includegraphics [scale=0.5,angle=0]{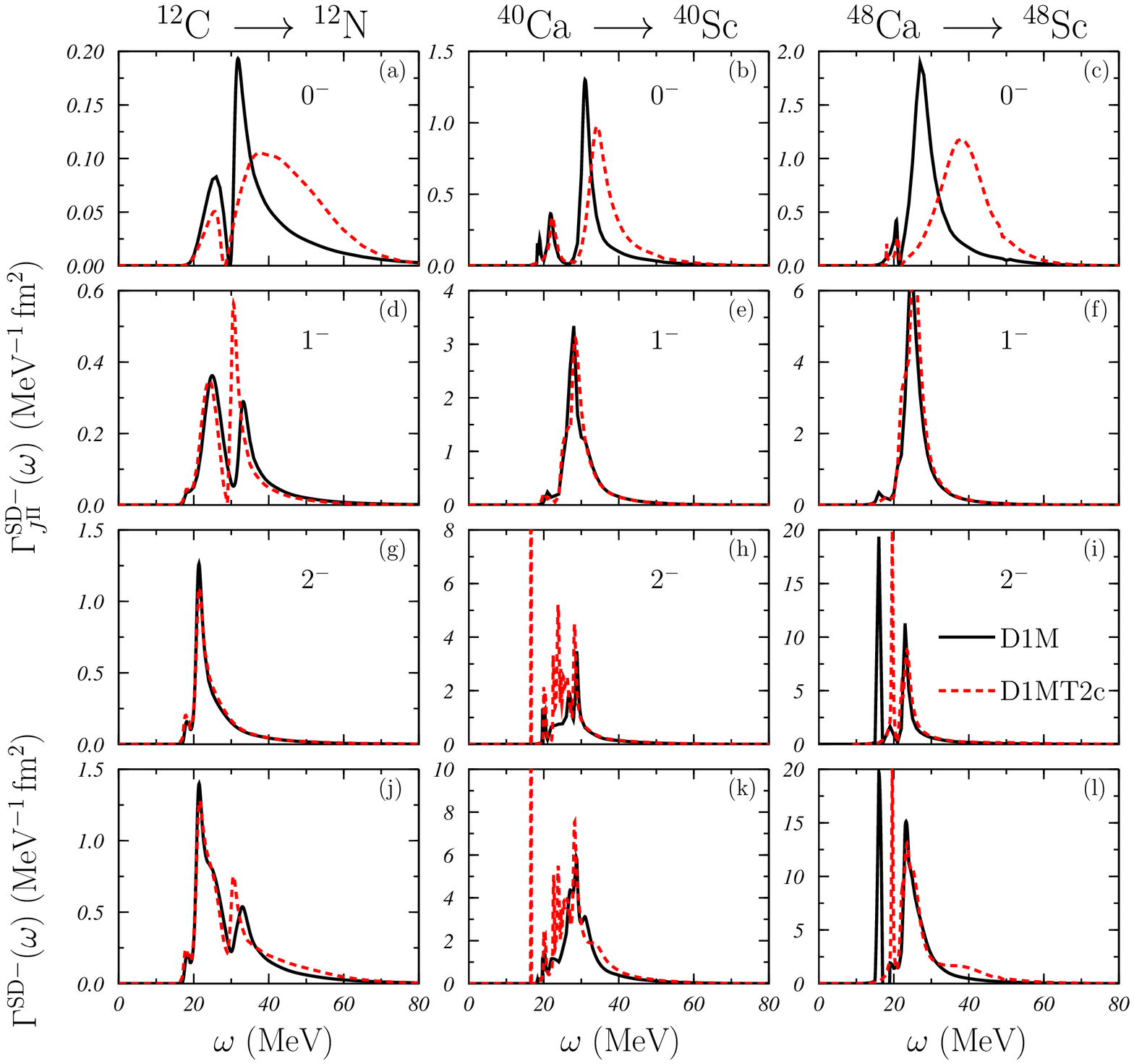} 
\vspace*{-0.3cm}
\caption{\small  The same as in Fig. \ref{fig:SD_oxy} for \carI, \caI, and \caII nuclei. 
} 
\label{fig:SD_C-Ca} 
\end{center} 
\end{figure}

\begin{figure}[!b] 
\begin{center} 
\includegraphics [scale=0.5,angle=0]{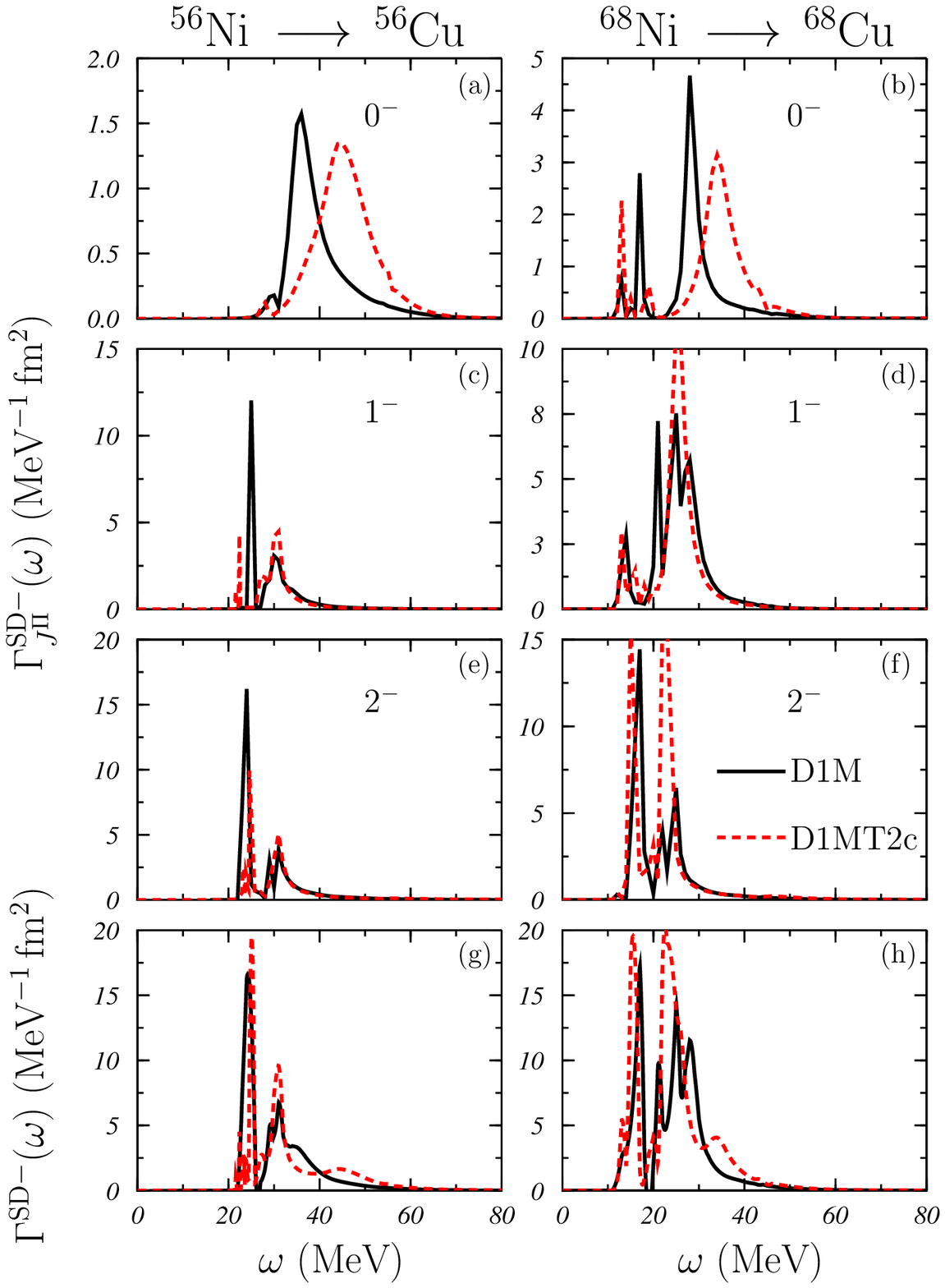} 
\vspace*{-0.3cm}
\caption{\small  The same as in Fig. \ref{fig:SD_oxy} for \nicI, and \nicII nuclei. 
} 
\label{fig:SD_nic} 
\end{center} 
\end{figure}

We show in Figs. \ref{fig:SD_oxy}, \ref{fig:SD_C-Ca}, and
\ref{fig:SD_nic} the \tm SD strength distributions for all the nuclei
we have considered. We present separately the contribution of each
multipole, and also the total strength. As in the previous figures,
the full black lines show the results obtained with the D1M
interaction, while the dashed red lines show those obtained with D1MT2c.
In these cases the size of the strength is comparable with that
obtained in DRPA calculations \cite{don14b}, since the main part of
the strength is above the particle emission threshold.  All the cases
investigated indicate that the largest contribution to the total
strength comes from the $2^-$ excitation, which is about one order of
magnitude larger than that of the $0^-$, with the only exception being the $^{68}$Ni nucleus,
where it is three times larger.

The inclusion of the tensor force has a small influence on the $1^-$
and $2^-$ strength distributions, while the changes on the $0^-$
strength, the smallest one, are remarkable. In all the $0^-$ cases
considered, the tensor force moves the peak of the resonance toward
higher energies.  In order to present a more quantitative information
of this effect, we show in Table \ref{tab:centens} the values of the
centroid energies and of the variances of the $0^-$ main resonances
shown in Figs.  \ref{fig:SD_oxy}, \ref{fig:SD_C-Ca} and
\ref{fig:SD_nic}.  In general, the shift of the centroid energies is
relevant and reaches a value of about $9\,$MeV, in the case of the
$^{22}$O, and $^{24}$O nuclei.  The effects on the variance,
indicating a larger spreading of the $0^-$ width, are significant only
in the oxygen isotopes.
  
%
%
\begin{figure}[ht] 
\begin{center} 
\includegraphics [scale=0.5,angle=0]{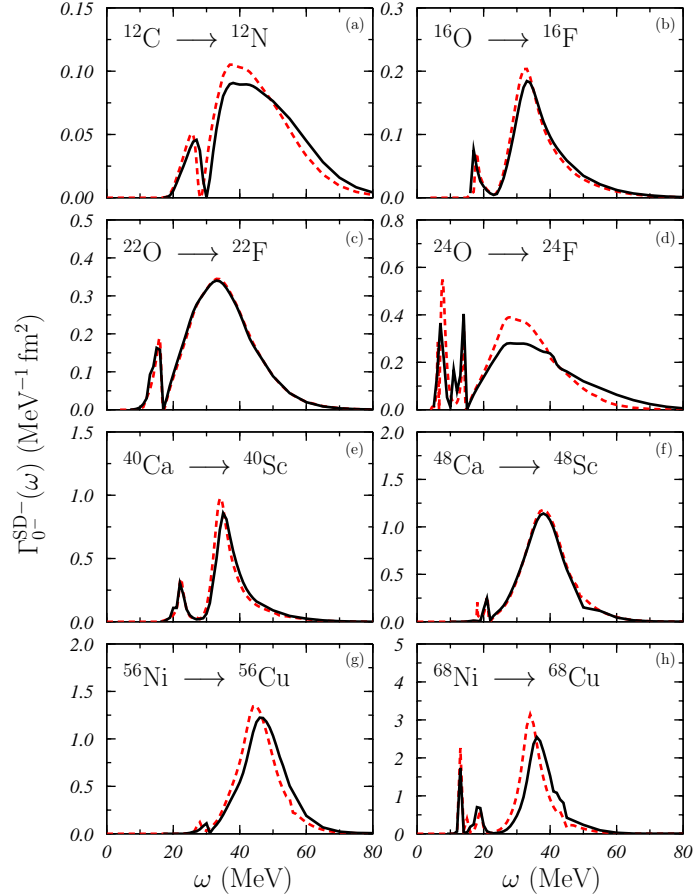} 
\vspace*{-0.3cm}
\caption{\small  Energy distributions of the \tm 
$\Gamma_{0^-}^{{\rm SD}}(\omega)$ obtained in CRPA calculations for the various nuclei studied.   
The dashed red curves correspond to the D1MT2c force. 
The solid black lines have been obtained by using 
the D1M s.p. wave functions together with the D1MT2c interaction in the CRPA calculations.
} 
\label{fig:SD0_D1M} 
\end{center} 
\end{figure}

In our calculations with the D1MT2c interaction, the tensor is active
in both HF and RPA calculations. In order to disentangle these two effects, in
Fig. \ref{fig:SD0_D1M}, we show the $0^-$ excitations for all the
nuclei we have investigated.  The dashed red curves have been obtained
by using the D1MT2c interaction in both HF and CRPA calculations, and
they coincide with those shown in Figs. \ref{fig:SD_oxy}, \ref{fig:SD_C-Ca},
and \ref{fig:SD_nic}. The solid black curves indicate the results
obtained in CRPA calculations by using the D1MT2c interaction but with
the HF s.p. wave functions generated with the D1M force.  The
comparison between these results and the full lines of the $0^-$
excitations of Figs. \ref{fig:SD_oxy}, \ref{fig:SD_C-Ca}, and
\ref{fig:SD_nic}, indicates that the effect of the tensor is mainly
present in the RPA, rather than in the modification of the s.p. wave
functions.

\begin{table}[htb] 
\begin{center} 
\begin{tabular}{cccccccc} 
\hline \hline 
& & & \multicolumn{2}{c}{${\omega}_{{\rm cen},0^-}^{{\rm SD}-}$ (MeV)} &~~~& \multicolumn{2}{c}{$\var_{0^-}^{{\rm SD}-}$ (MeV)} \\ \cline{4-5} \cline{7-8}
nucleus&  $E_<$ (MeV) & $E_>$ (MeV) & D1M & D1MT2c && D1M & D1MT2c \\  
\hline 
\oxyI  & 20.0 & 60.0 & 32.82 & 36.45 && 7.00 & 7.34  \\ 
\oxyII & 16.7 & 60.0 & 26.35 & 35.28 && 7.57 & 8.67  \\ 
\oxyIII& 15.0 & 60.0 & 24.18 & 33.44 && 7.68 & 9.15  \\ 
\carI  & 29.6 & 60.0 & 38.67 & 43.62 && 7.38 & 7.64  \\ 
\caI   & 26.0 & 60.0 & 34.32 & 37.57 && 5.68 & 5.80  \\ 
\caII  & 21.4 & 60.0 & 30.63 & 39.12 && 6.42 & 6.79  \\ 
\nicI  & 30.0 & 60.0 & 39.43 & 45.36 && 5.86 & 5.69  \\ 
\nicII & 20.0 & 50.0 & 30.33 & 35.24 && 4.62 & 4.60  \\ 
\hline 
\hline 
\end{tabular} 
\vspace*{-0.2cm}
\caption{ Values of the centroid energies and variances, as given by Eqs. (\ref{eq:centr-mul}) and (\ref{eq:var}), respectively, for the T$_-$ SD,  $J^\Pi=0^-$ multipole excitations, for the D1M and D1MT2c interactions. We have indicated with $E_<$ 
  and $E_>$, in MeV, the extremes of the energy integral in
  Eq. \eqref{eq:momJ}.   
} 
\label{tab:centens} 
\end{center} 
\end{table} 

A crucial test of the validity of the theoretical results is to check
if the sum rules, obtained for the F, GT, and SD excitations by
evaluating Eq. \eqref{eq:exha} for large values of the excitation energy,
are satisfied according to the expected values. In
Ref. \cite{don14b}, we have shown that our DRPA self-consistent
approach satisfies the sum rules within 0.1\%. An analogous test
with the CRPA calculations is more difficult, since a large part of the
strength is distributed below the emission particle threshold, where
this approach is not applicable.  For this reason, we have added to
the CRPA sum-rule exhaustion functions \eqref{eq:exha} the
contribution of the DRPA below threshold and compared the result with the DRPA
asymptotic values.

%
%
\begin{figure}[ht] 
\begin{center} 
\includegraphics [scale=0.5,angle=0]{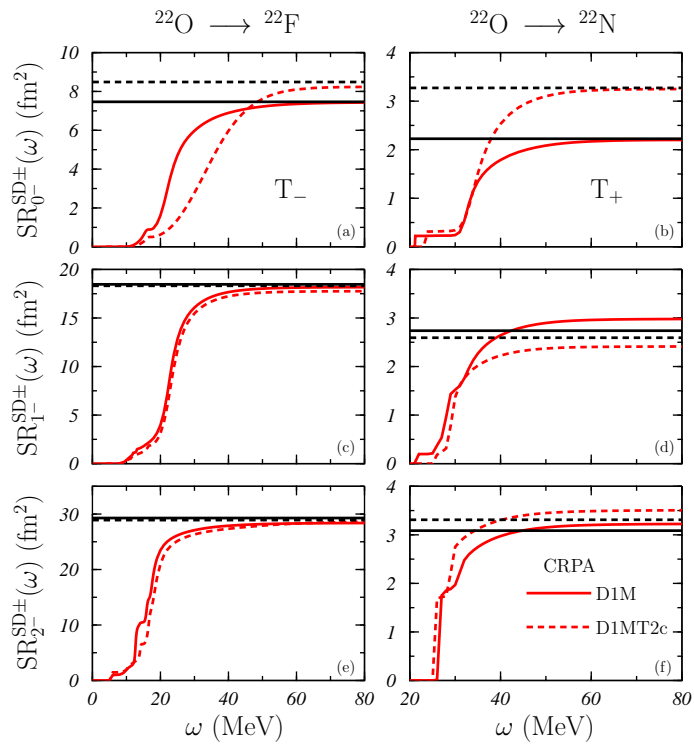} 
\vspace*{-0.3cm}
\caption{\small SD sum rule exhaustion, as given by 
Eq. (\ref{eq:exha}), obtained in CRPA calculations  
for the $^{22}$O nucleus. 
Left (right) panels correspond to the T$_-$ (T$_+$) transitions.    
The full red curves show the results obtained with the D1M force
and the dashed red curves those obtained with the D1MT2c force.  
The full and dashed horizontal black lines indicated the asymptotic
values of the analogous DRPA results taken from 
 Ref. \cite{don14b}.
} 
\label{fig:SR_all} 
\end{center} 
\end{figure} 
As example of this procedure, we show in Fig. \ref{fig:SR_all} the
results obtained with the three multipole excitations of the SD
operator for the \oxyII nucleus.  The curves show the CRPA sum rule
exhaustion functions \eqref{eq:exha} as a function of $\omega$, the
upper limit of the integral.  The full red curve was obtained
with the D1M interaction while the dashed red curve was obtained with the D1MT2c
interaction. The horizontal black lines indicate the asymptotic values
we have obtained with the analogous DRPA calculations.

The agreement between the results of the two calculations is
remarkable, even though we observe a slight overshooting in the \tp
responses for $1^-$ and $2^-$.  For the $0^-$ multipolarity, the
asymptotic value is reached more quickly in the D1M case than in the
D1MT2c one. This is due to the shift of the peak of the resonance
towards larger energy and to the bigger spreading of the strengths
when tensor terms of the interactions are active, as we have already
stressed and shown in Fig.~\ref{fig:SD_oxy} .  It is remarkable the
increase of the \tm and \tp strengths generated by the tensor term. We
recall that the theoretical sum rule asymptotic value limiting the
strength is obtained as a difference between the \tm and \tp
strengths, which is conserved in all the $0^-$ calculations, with or
without tensor force.
 
\section{Summary and conclusions}    
\label{sec:summary} 
We have extended the CRPA formalism of Ref. \cite{don11a}, which
handles finite-range interactions and continuum s.p. configuration
space without any approximation, to the description of charge-exchange
excitations.  In our calculations we used four Gogny-like finite-range
interactions, two of them containing tensor terms, and we applied our
formalism to describe the F, GT, and SD excitations in the $^{12}$C,
$^{16}$O, $^{22}$O, $^{24}$O, $^{40}$Ca, $^{48}$Ca, $^{56}$Ni, and
$^{68}$Ni nuclei. The goals of our work were:
\begin{enumerate}
\item the presentation of  the non trivial extension of the 
charge-conserving CRPA formalism to treat charge-exchange excitations;
\item the comparison of CRPA results with those obtained 
in DRPA calculations;
\item the study of the effects of the tensor interaction
in charge-exchange excitations, and
\item the comparison with the available experimental data.
\end{enumerate}

Concerning the first goal, we have shown that it is possible to
formulate the $pn$-CRPA secular equations in a form which allows us to
obtain simultaneous solution of proton-neutron and neutron-proton
excitations. This is analogous to the commonly used formulation of the
$pn$-DRPA \cite{hal67}, as it has been used, for example, in Ref.
\cite{don14b}.

The comparison between CRPA and DRPA results has been carried out to
test the stability of the DRPA calculations against the dimensions of
the s.p. configuration space.  We have obtained a very good agreement
between the position of the DRPA peaks and that of the maxima of the
CRPA strength. For a given multipole excitation, by using a folding
procedure with a Lorentz function of appropriate width, we obtained
agreement also with the CRPA strength distribution. Furthermore, we
have also verified that the two types of calculations satisfy the sum
rules with the same degree of accuracy.

We have pointed out the ambiguities of the folding procedure commonly
used to generate continuous responses out of DRPA solutions. 
We have shown that the three CRPA components of the SD
response cannot be simultaneously described by using a single 
Lorentz function.
Having established the agreement between CRPA and DRPA results, we are
confident on the use of DRPA calculations in situations where we
cannot carry out CRPA calculations. These are, for example, the
cases of excitations below the particle emission threshold, and the
calculations for nuclei heavier than those treated in the present
article. 

Our formalism allows the use of effective nucleon-nucleon interactions
of finite-range type which include tensor terms.  In
Ref. \cite{don14b} we presented charge-exchange results obtained with
the same interactions used here. The results shown in this reference
have been obtained within the DRPA framework.  The present work can be
considered to complement that of Ref.  \cite{don14b} by considering
the continuum.  

The tensor force does not produce remarkable effects on the continuum
part of the F and GT strengths, which, for nuclei with neutron excess,
is only a small part of the total strength.  The situation is
different for the SD excitations whose strength develops mainly in the
continuum. We have analyzed the $0^-$, $1^-$, and $2^-$ multipoles
composing the SD responses for various nuclei and we found that the
$2^-$ excitation is the most important of the three, dominating the
total response.  On the other hand, the $0^-$ strength is extremely
sensitive to the tensor term. Its inclusion in the CRPA calculations
moves the maxima of the $0^-$ responses towards higher energies.  This
shift is particularly remarkable in the oxygen isotopes where we found
changes of the centroid energies of up to 9 MeV, and a significant
widening of the width. We have verified that this large sensitivity of
the $0^-$ to the tensor force is mainly related to the RPA
calculations, and not to the modifications of the s.p. energies.

The fourth, and last, goal of our work was the comparison with the
available data, which we have found mainly for the GT transitions.  In
our approach, which considers only 1p-1h excitations, the GT
excitations are strongly hindered in nuclei with equal number of
protons and neutrons and with spin-orbit partner levels fully
occupied.  The experimental strength of \oxyI \cite{mer94} is one
order of magnitude larger than that of our CRPA
calculations. Also the experimental spectrum of $^{16}$F obtained
  in charge-exchange experiments on the \oxyI target
  \cite{faz82,mer94,mad97,fuj09} is not described in detail by our
  approach, even though the order of magnitude of the excitation
  energies is reproduced. These are indications of the need to
  include higher order particle-hole excitations to improve the
  description of these data.

The comparison with the GT data in \caII and \nicI shows the need to include
the spreading width to obtain a detailed description of the
experiment. On the other hand, the position of the main peaks of the
resonance is rather well reproduced. 
The experimentally
measured strengths in these two nuclei satisfy only part of the sum
rule. Our calculations indicate that the GT strength above the
measured highest excitation energy is not negligible. This finding is
in agreement with the study of $^{90}$Zr (p,n) $^{90}$Nb data
\cite{ich06}, where, by considering the strength lying beyond the main
resonance, all the expected GT strength has been found.

Despite the limitations we have pointed out, our CRPA model, which
considers all the continuum s.p. space without the need of artificial
cuts, is a further step towards the construction of a parameter free
mean-field approach describing nuclei in all the regions of the
nuclear chart. We do not aim for a detailed description of
  low-lying spectra and strength distributions; however peak
positions and global strengths can be well predicted by our approach,
which can be applied also to unstable nuclei.  
This offers a great potential for the study of neutrino cross
sections in the energy region around the
emission particle threshold, in analogy to what has been done in
Ref. \cite{bot05} for stable nuclei.
 
%
\appendix
\section{$pn$-CRPA matrix elements.} 
\label{sec:appmat} 

In this Appendix we obtain the $pn$-CRPA equations. First we recall
here the basic equations of our CRPA formalism. A more detailed
description can be found in Refs. \cite{don08t,don09,don11a}.

For each excited state $|\alpha \rangle \equiv |J,\Pi,\omega \rangle$, 
characterized by its total angular momentum $J$, parity $\Pi$ and 
excitation energy $\omega$, we write a set of algebraic equations
whose unknowns are the $c^\mu_{ph}$ expansion coefficients 
\cite{don08t,don09}
\beqn 
\hspace*{-0.5cm}  
\sum_\mu \, \sum_{p'h'} \, \Bigg\{ \Bigg[ \delta_{pp'} \, \delta_{hh'} 
\Big( \delta_{\mu \alpha} \, - 
\, \langle (\Phi_p^{\alpha +})^*|\cau|\Phi_p^{\mu+} \rangle \,+ \,  
\langle (\Phi_p^{\alpha +})^*\, \ide |\caw| \ide \,\Phi^{\mu+}_p \rangle \nonumber \\ 
&& \hspace*{-7.6cm} + \, \sum_{\epsilon_i<\epsilon_{\rm F}}\, \delta_{ip}\, (\epsilon_i-\epsilon_h-\omega)\, 
\langle (\Phi_p^{\alpha +})^* |R_i\rangle \langle 
(R_i)^*|\Phi_p^{\mu+}\rangle \Big)\nonumber \\ 
&& \hspace*{-10.cm}  - \, \Big( 
\langle (\widetilde{\Phi}_p^{\alpha +})^* R_{h'} |   
V^{J,{\rm dir}}_{ph,p'h'} |R_h \widetilde{\Phi}^{\mu+}_{p'} \rangle \, 
- \, \langle (\widetilde{\Phi}_p^{\alpha +})^* R_{h'} | 
   V^{J,{\rm exc}}_{ph,p'h'} | \widetilde{\Phi}^{\mu+}_{p'} R_h  \rangle \Big) 
\Bigg] \, c^{\mu+}_{p'h'} \, \nonumber\\ 
&& \hspace*{-10.cm}  - \, \Big( 
\langle (\widetilde{\Phi}_p^{\alpha +})^*  \widetilde{\Phi}^{\mu-}_{p'}|   
U^{J,{\rm dir}}_{ph,p'h'} | R_h R_{h'} \rangle \,  
- \, \langle (\widetilde{\Phi}_p^{\alpha +})^*  \widetilde{\Phi}^{\mu-}_{p'}|   
U^{J,{\rm exc}}_{ph,p'h'} | R_{h'} R_h \rangle \Big) \, (c^{\mu-}_{p'h'})^* 
\Bigg\} \, = \, \nonumber \\ 
&& \hspace*{-10.5cm} 
= \, \langle (\widetilde{\Phi}_p^{\alpha +})^* R_{h_0} | 
   V^{J,{\rm dir}}_{ph,p_0h_0} |R_h R_{p_0}(\epsilon_{p_0}) \rangle \,  
- \, \langle (\widetilde{\Phi}_p^{\alpha +})^* R_{h_0} | 
   V^{J,{\rm exc}}_{ph,p_0h_0} |R_{p_0}(\epsilon_{p_0}) R_h  \rangle  
\label{eq:st1}  
\eeqn 
and
\beqn 
\hspace*{-0.5cm}  
\sum_\mu \, \sum_{p'h'} \, \Bigg\{ \Bigg[ \delta_{pp'} \, \delta_{hh'} 
\Big( \delta_{\mu \alpha} \, - 
\, \langle (\Phi_p^{\alpha -})^*|\cau|\Phi_p^{\mu-} \rangle \,+ \,  
\langle (\Phi_p^{\alpha -})^*\, \ide |\caw| \ide \,\Phi^{\mu-}_p \rangle \nonumber \\ 
&& \hspace*{-7.6cm} + \, \sum_{\epsilon_i<\epsilon_{\rm F}}\, \delta_{ip}\, (\epsilon_i-\epsilon_h+\omega)\, 
\langle (\Phi_p^{\alpha -})^* |R_i\rangle \langle 
(R_i)^*|\Phi_p^{\mu-}\rangle \Big)\nonumber \\ 
&& \hspace*{-10.cm}  - \, \Big( 
\langle (\widetilde{\Phi}_p^{\alpha -})^* R_{h'} |   
V^{J,{\rm dir}}_{ph,p'h'} |R_h \widetilde{\Phi}^{\mu-}_{p'} \rangle \, 
- \, \langle (\widetilde{\Phi}_p^{\alpha -})^* R_{h'} | 
   V^{J,{\rm exc}}_{ph,p'h'} | \widetilde{\Phi}^{\mu-}_{p'} R_h  \rangle \Big) 
\Bigg] \, c^{\mu-}_{p'h'} \, \nonumber\\ 
&& \hspace*{-10.cm}  - \, \Big( 
\langle (\widetilde{\Phi}_p^{\alpha -})^*  \widetilde{\Phi}^{\mu+}_{p'}|   
U^{J,{\rm dir}}_{ph,p'h'} | R_h R_{h'} \rangle \,  
- \, \langle (\widetilde{\Phi}_p^{\alpha -})^*  \widetilde{\Phi}^{\mu+}_{p'}|   
U^{J,{\rm exc}}_{ph,p'h'} | R_{h'} R_h \rangle \Big) \, (c^{\mu+}_{p'h'})^* 
\Bigg\} \, = \, \nonumber \\ 
&& \hspace*{-10.5cm} 
= \, \langle (\widetilde{\Phi}_p^{\alpha -})^* R_{p_0}(\epsilon_{p_0}) | 
   U^{J,{\rm dir}}_{ph,p_0h_0} |R_h R_{h_0} \rangle \,  
- \, \langle (\widetilde{\Phi}_p^{\alpha -})^* R_{p_0}(\epsilon_{p_0}) | 
   U^{J,{\rm exc}}_{ph,p_0h_0} |R_{h_0} R_h  \rangle \, . 
\label{eq:st2}  
\eeqn 
In the above equations we have indicated, respectively, as 
$\cau$ and $\caw$ the Hartree and Fock-Dirac potentials as 
they are commonly defined in the HF formalism \cite{rin80}.
The symbols  $U$ and $V$ indicate the 
matrix elements of the nucleon-nucleon interaction,
$R_h$ the s.p. radial wave function of a hole state of 
$\epsilon_h$ energy, and $R_p(\epsilon_p)$ 
the s.p. radial wave function of a continuum particle state.

We simplify the writing of Eqs. \eqref{eq:st1} and \eqref{eq:st2} by defining 
the following quantities:
\beqn 
A^{\mu \pm}_{ph,p'h'}&=&\delta_{pp'} \, \delta_{hh'} 
\Big( \delta_{\mu \alpha} \, - 
\, \langle (\Phi_p^{\alpha \pm})^*|\cau|\Phi_p^{\mu \pm} \rangle \,+ \,  
\langle (\Phi_p^{\alpha \pm})^*\, \ide |\caw| \ide \,\Phi^{\mu \pm}_p \rangle  
\nonumber \\ 
&& \hspace*{2.3cm} + \, \sum_{\epsilon_i<\epsilon_{\rm F}}\, \delta_{ip}\, ( 
\epsilon_i-\epsilon_h\mp \omega)\, 
\langle (\Phi_p^{\alpha \pm})^* |R_i\rangle \langle 
(R_i)^*|\Phi_p^{\mu\pm}\rangle \Big) \nonumber \\ 
&& \hspace*{0.1cm} - \, \left(
\langle (\widetilde{\Phi}_p^{\alpha \pm})^* R_{h'} |   
V^{J,{\rm dir}}_{ph,p'h'} |R_h \widetilde{\Phi}^{\mu\pm}_{p'} \rangle \, 
- \, \langle (\widetilde{\Phi}_p^{\alpha \pm})^* R_{h'} | 
   V^{J,{\rm exc}}_{ph,p'h'} | \widetilde{\Phi}^{\mu\pm}_{p'} R_h  \rangle  \right)
 \\ 
B^{\mu \pm}_{ph,p'h'}&=& 
\langle (\widetilde{\Phi}_p^{\alpha \pm})^*  \widetilde{\Phi}^{\mu \mp}_{p'}|   
U^{J,{\rm dir}}_{ph,p'h'} | R_h R_{h'} \rangle \,  
- \, \langle (\widetilde{\Phi}_p^{\alpha \pm})^*  \widetilde{\Phi}^{\mu \mp}_{p'}|   
U^{J,{\rm exc}}_{ph,p'h'} | R_{h'} R_h \rangle \\ 
C_{ph,p_0h_0}&=&\langle (\widetilde{\Phi}_p^{\alpha +})^* R_{h_0} | 
   V^{J,{\rm dir}}_{ph,p_0h_0} |R_h R_{p_0}(\epsilon_{p_0}) \rangle \,  
- \, \langle (\widetilde{\Phi}_p^{\alpha +})^* R_{h_0} | 
   V^{J,{\rm exc}}_{ph,p_0h_0} |R_{p_0}(\epsilon_{p_0}) R_h \rangle \\
D_{ph,p_0h_0}&=&\langle (\widetilde{\Phi}_p^{\alpha -})^* R_{p_0}(\epsilon_{p_0}) | 
   U^{J,{\rm dir}}_{ph,p_0h_0} |R_h R_{h_0} \rangle \,  
- \, \langle (\widetilde{\Phi}_p^{\alpha -})^* R_{p_0}(\epsilon_{p_0}) | 
   U^{J,{\rm exc}}_{ph,p_0h_0} |R_{h_0} R_h  \rangle 
   \, ,
\eeqn 
and we obtain
\beqn 
\begin{bmatrix} 
A^{\mu +}_{ph,p'h'} & -B^{\mu +}_{ph,p'h'} \\  
-(B^{\mu -}_{ph,p'h'})^* & (A^{\mu -}_{ph,p'h'})^*  \\  
\end{bmatrix} 
\begin{bmatrix} 
c^{\mu +}_{p'h'}\\  
(c^{\mu -}_{p'h'})^* \\ 
\end{bmatrix}= 
\begin{bmatrix} 
C_{ph,p_0h_0}\\ 
(D_{ph,p_0h_0})^*\\ 
\end{bmatrix}  
\label{eq:CRPA2} 
\, ,
\eeqn 
where the sums over $\mu$ and $(p'h')$ are understood.

In the case of charge-exchange excitations, the $ph$ pairs may be either $\pi\bn$, 
for $T_-$, or $\nu\bpp$, for $T_+$. The extended expression of Eq. \eqref{eq:CRPA2} 
is:
\beqn 
\begin{bmatrix} 
A^{\mu +}_{\pi \bn,\pi' \bn'} & A^{\mu +}_{\pi\bn,\nu' \bpp'} & -B^{\mu +}_{\pi\bn,\pi' \bn'} & -B^{\mu +}_{\pi\bn,\nu' \bpp'} \\  
A^{\mu +}_{\nu\bpp,\pi' \bn'} & A^{\mu +}_{\nu\bpp,\nu' \bpp'} & -B^{\mu +}_{\nu\bpp,\pi' \bn'} & -B^{\mu +}_{\nu\bpp,\nu' \bpp'} \\  
-(B^{\mu -}_{\pi\bn,\pi' \bn'})^* & -(B^{\mu -}_{\pi\bn,\nu' \bpp'})^* & (A^{\mu -}_{\pi\bn,\pi' \bn'})^* & (A^{\mu -}_{\pi\bn,\nu' \bpp'})^* \\  
-(B^{\mu -}_{\nu\bpp,\pi' \bn'})^* & -(B^{\mu -}_{\nu\bpp,\nu' \bpp'})^* & (A^{\mu -}_{\nu\bpp,\pi' \bn'})^* & (A^{\mu -}_{\nu\bpp,\nu' \bpp'})^*  \\  
\end{bmatrix} 
\begin{bmatrix} 
c^{\mu +}_{\pi' \bn'}\\  
c^{\mu +}_{\nu' \bpp'}\\  
(c^{\mu -}_{\pi' \bn'})^* \\ 
(c^{\mu -}_{\nu' \bpp'})^* \\ 
\end{bmatrix}= 
\begin{bmatrix} 
C_{\pi \bn,p_0h_0}\\ 
C_{\nu \bpp,p_0h_0}\\ 
(D_{\pi \bn,p_0h_0})^*\\ 
(D_{\nu \bpp,p_0h_0})^*\\ 
\end{bmatrix}  
\label{eq:pnCRPA} 
\, .
\eeqn 
The requirement of charge-conservation implies
\beqn
A^{\mu \pm}_{\pi\bn,\nu' \bpp'} \, = \, B^{\mu \pm}_{\pi\bn,\pi' \bn'} \, = \, A^{\mu \pm}_{\nu\bpp,\pi' \bn'} \, = \, B^{\mu \pm}_{\nu\bpp,\nu' \bpp'}  \, = \,  0 \, ,
\eeqn
therefore Eq. (\ref{eq:pnCRPA}) reduces to
\beqn 
\begin{bmatrix} 
A^{\mu +}_{\pi \bn,\pi' \bn'} & 0 & 0 & -B^{\mu +}_{\pi\bn,\nu' \bpp'} \\  
0 & A^{\mu +}_{\nu\bpp,\nu' \bpp'} & -B^{\mu +}_{\nu\bpp,\pi' \bn'} & 0 \\  
0 & -(B^{\mu -}_{\pi\bn,\nu' \bpp'})^*& (A^{\mu -}_{\pi\bn,\pi' \bn'})^* & 0 \\  
-(B^{\mu -}_{\nu\bpp,\pi' \bn'})^* & 0 & 0 & (A^{\mu -}_{\nu\bpp,\nu' \bpp'})^*  \\  
\end{bmatrix} 
\begin{bmatrix} 
c^{\mu +}_{\pi' \bn'}\\  
c^{\mu +}_{\nu' \bpp'}\\  
(c^{\mu -}_{\pi' \bn'})^* \\ 
(c^{\mu -}_{\nu' \bpp'})^* \\ 
\end{bmatrix}= 
\begin{bmatrix} 
C_{\pi \bn,p_0h_0}\\ 
C_{\nu \bpp,p_0h_0}\\ 
(D_{\pi \bn,p_0h_0})^*\\ 
(D_{\nu \bpp,p_0h_0})^*\\ 
\end{bmatrix}  
\label{eq:pnCRPA00} 
\, .
\eeqn 

This equation can be separated in two matrix equations:
\beqn 
\begin{bmatrix} 
A^{\mu +}_{\pi \bn,\pi' \bn'} & -B^{\mu +}_{\pi\bn,\nu' \bpp'} \\  
-(B^{\mu -}_{\nu\bpp,\pi' \bn'})^* & (A^{\mu -}_{\nu\bpp,\nu' \bpp'})^* \\  
\end{bmatrix} 
\begin{bmatrix} 
c^{\mu +}_{\pi' \bn'}\\  
(c^{\mu -}_{\nu' \bpp'})^* \\ 
\end{bmatrix}= 
\begin{bmatrix} 
C_{\pi \bn,p_0h_0}\\ 
(D_{\nu \bpp,p_0h_0})^*\\ 
\end{bmatrix}  
\label{eq:pnCRPA01} 
\eeqn 
and
\beqn 
\begin{bmatrix} 
A^{\mu +}_{\nu\bpp,\nu' \bpp'} & -B^{\mu +}_{\nu\bpp,\pi' \bn'}  \\  
-(B^{\mu -}_{\pi\bn,\nu' \bpp'})^* & (A^{\mu -}_{\pi\bn,\pi' \bn'})^* \\  
\end{bmatrix} 
\begin{bmatrix} 
c^{\mu +}_{\nu' \bpp'}\\  
(c^{\mu -}_{\pi' \bn'})^* \\ 
\end{bmatrix}= 
\begin{bmatrix} 
C_{\nu \bpp,p_0h_0}\\ 
(D_{\pi \bn,p_0h_0})^*\\ 
\end{bmatrix}  
\label{eq:pnCRPA02} 
\, .
\eeqn 

For $T_-$ transitions, the elastic channel is  $p_0h_0\equiv \pi_0 \bn_0$ 
and $c^{\mu +}_{\nu' \bpp'}=c^{\mu -}_{\pi' \bn'}=0$, therefore 
only Eq. \eqref{eq:pnCRPA01} survives. For the $T_+$ transition, we have 
$p_0h_0\equiv \nu_0 \bpp_0$ and 
$c^{\mu +}_{\pi' \bn'}=c^{\mu -}_{\nu' \bpp'}=0$, therefore only Eq. 
\eqref{eq:pnCRPA02} must be considered.

\acknowledgments  
This work has been partially supported by  
the Junta de Andaluc\'{\i}a (FQM0220), the European 
Regional Development Fund (ERDF), and the Spanish Ministerio de 
Econom\'{\i}a y Competitividad (FPA2012-31993). 
 

\begin{thebibliography}{10}
\expandafter\ifx\csname url\endcsname\relax
  \def\url#1{\texttt{#1}}\fi
\expandafter\ifx\csname urlprefix\endcsname\relax\def\urlprefix{URL }\fi
\expandafter\ifx\csname href\endcsname\relax
  \def\href#1#2{#2} \def\path#1{#1}\fi

\bibitem{arn07}
M.~Arnould, S.~Goriely, and K.~Takahashi, Phys. \ Rep. 450 (2007) 97.

\bibitem{boh52}
D.~Bohm and D.~Pines, Phys. \ Rev. 92 (1952) 609.

\bibitem{row70}
D.~J. Rowe, Nuclear Collective Motion, Methuen, London, 1970.

\bibitem{hal67}
J.~A. Halbleib and R.~A. Sorensen, Nucl. \ Phys. \ A 98 (1967) 542.

\bibitem{lan80}
A.~M. Lane and J.~Martorell, Ann. \ Phys. \ (N.Y.) 129 (1980) 273.

\bibitem{aue81}
N.~Auerbach, A.~Klein, and N.~V. Giai, Phys. \ Lett. \ B 106 (1981) 347.

\bibitem{aue83}
N.~Auerbach and A.~Klein, Nucl. \ Phys. \ A 395 (1983) 77.

\bibitem{mig67}
A.~Migdal, Theory of Finite Fermi Systems and Applications to Atomic Nuclei
  (Interscience, London, 1967).

\bibitem{ost92}
F.~Osterfeld, Rev. \ Mod. \ Phys. 64 (1992) 491.

\bibitem{har01}
M.~N. Harakeh and A.~van~der Woude, Giant Resonances (Clarendon press, Oxford,
  2001).

\bibitem{ich06}
M.~Ichimura, H.~Sakai, and T.~Wakasa, Prog. \ Part. \ Nucl. \ Phys. 56 (2006) 446.

\bibitem{cha07t}
F.~Chappert, Ph.D. thesis, Universit\'e de Paris-Sud XI (France), 2007,
  http://tel.archives-ouvertes.fr/tel-001777379/en/.

\bibitem{ham00}
I.~Hamamoto and H.~Sagawa, Phys. \ Rev. \ C 62 (2000) 024319.

\bibitem{fra07}
S.~Fracasso and G.~Col\`o, Phys. \ Rev. \ C 76 (2007) 044307.

\bibitem{bai09a}
C.~L. Bai, H.~Sagawa, H.~Q. Zhang, X.~Z. Zhang, G.~Col\`o, and F.~R. Xu, Phys. \
  Lett. \ B 675 (2009) 28.

\bibitem{bai09b}
C.~L. Bai, H.~Q. Zhang, X.~Z. Zhang, F.~R. Xu, H.~Sagawa, and G.~Col\`o, Phys. \
  Rev. \ C 79 (2009) 041301(R).

\bibitem{bai10}
C.~L. Bai, H.~Q. Zhang, H.~Sagawa, X.~Z. Zhang, G.~Col\`o, and F.~R. Xu, Phys. \
  Rev. \ Lett. 105 (2010) 072501.

\bibitem{bai11a}
C. L. Bai, H. Q. Zhang, H. Sagawa, X. Z. Zhang, G. Col\`o, and F. R. Xu, Phys. \ Rev. \ C 83 (2011) 054316.

\bibitem{bai11b}
C.~L. Bai, H.~Sagawa, G.~Col\`o, H.~Q. Zhang, and X.~Z. Zhang, Phys. \ Rev. \ C 84
  (2011) 044329.

\bibitem{min13}
F.~Minato and C.~L. Bai, Phys. \ Rev. \ Lett. 110 (2013) 122501.

\bibitem{dec80}
J.~Decharg\`e and D.~Gogny, Phys. \ Rev. \ C 21 (1980) 1568.

\bibitem{don14b}
V.~{De Donno}, G.~Co', M.~Anguiano, and A.~M. Lallena, Phys. \ Rev. \ C 90 (2014)
  024326.

\bibitem{don11a}
V.~{De Donno}, G.~Co', M.~Anguiano, and A.~M. Lallena, Phys. \ Rev. \ C 83 (2011)
  044324.

\bibitem{don08t}
V.~{De Donno}, Ph.D. thesis, Universit\`a del Salento (Italy), 2008,
  http://www.fisica.unisalento.it/~gpco/stud.html.

\bibitem{don09}
V.~{De Donno}, G.~Co', C.~Maieron, M.~Anguiano, A.~M. Lallena, and
  M.~Moreno-Torres, Phys. \ Rev. \ C 79 (2009) 044311.

\bibitem{don11b}
V.~{De Donno}, M.~Anguiano, G.~Co', and A.~M. Lallena, Phys. \ Rev. \ C 84 (2011)
  037306.

\bibitem{co13}
G.~Co', V.~{De Donno}, M.~Anguiano, and A.~M. Lallena, Phys. Rev. C 87 (2013)
  034305.

\bibitem{rin80}
P.~Ring and P.~Schuck, The Nuclear Many-Body Problem (Springer, Berlin, 1980).

\bibitem{raw82}
G.~Rawitscher, Phys. \ Rev. \ C. 25 (1982) 2196.

\bibitem{edm57}
A.~R. Edmonds, Angular Momentum in Quantum Mechanics (Princeton University
  Press, Princeton, 1957).

\bibitem{gor09}
S.~Goriely, S.~Hilaire, M.~Girod, and S.~P\'eru, Phys. \ Rev. \ Lett. 102 (2009)
  242501.

\bibitem{ber91}
J.~F. Berger, M.~Girod, and D.~Gogny, Comp. \ Phys. \ Commun. 63 (1991) 365.

\bibitem{don14a}
V.~{De Donno}, G.~Co', M.~Anguiano, and A.~M. Lallena, Phys. \ Rev. \ C 89 (2014)
  014309.

\bibitem{co98b}
G.~Co' and A.~M. Lallena, Nuovo \ Cimento \ A 111 (1998) 527.

\bibitem{bau99}
A.~R. Bautista, G.~Co', and A.~M. Lallena, Nuovo \ Cimento \ A 112 (1999) 1117.

\bibitem{bnlw}
{Brookhaven National Laboratory}, National Nuclear Data Center,
  http://www.nndc.bnl.gov/.

\bibitem{aud03}
G.~Audi, A.~H. Wapstra, and C.~Thibault, Nucl. \ Phys. \ A 729 (2003) 337.

\bibitem{co12b}
G.~Co', V.~{De Donno}, M.~Anguiano, and A.~M. Lallena, Phys. \ Rev. \ C 85 (2012)
  034323.

\bibitem{mer94}
D.~J. Mercer, et~al., Phys. \ Rev. \ C 49 (1994) 3104.

\bibitem{bro94}
B.~A. Brown, Nucl. \ Phys. \ A 577 (1994) 13c.

\bibitem{faz82}
A.~Fazely, et~al., Phys. \ Rev. \ C 25 (1982) 1760.

\bibitem{mad97}
R.~Madey, et~al., Phys. \ Rev. \ C 56 (1997) 3210.

\bibitem{fuj09}
H.~Fujita, et~al., Phys. \ Rev. \ C 79 (2009) 024314.

\bibitem{yak09}
K.~Yako, et~al., Phys. \ Rev. \ Lett. 103 (1994) 012503.

\bibitem{spe91}
J.~Speth and J.~Wambach, in {\it Electric and Magnetic
  Giant Resonances in Nuclei}, edited by J. Speth (World Scientific, Singapore,
  1991).

\bibitem{gam12}
D.~Gambacurta, M.~Grasso, V.~{De Donno}, G.~Co', and F.~Catara, Phys.\ Rev. \ C 86
  (2012) 021304(R).

\bibitem{ots05}
T.~Otsuka, T.~Suzuki, R.~Fujimoto, H.~Grawe, and Y.~Akaishi, Phys. \ Rev. \ Lett.
  95 (2005) 232502.

\bibitem{sas11}
M.~Sasano, et~al., Phys. \ Rev. \ Lett. 107 (2011) 202501.

\bibitem{bai13}
C.~L. Bai, et~al., Phys. \ Lett. \ B 719 (2013) 116.

\bibitem{bot05}
A.~Botrugno and G.~Co', Nucl. \ Phys. \ A 761 (2005) 203.

\end{thebibliography}
%

\end{document}